\documentclass{article}

\usepackage {amsmath,amssymb,latexsym}
\usepackage {delarray,graphics,graphpap,alltt,graphicx}
\def\varphi{\Phi}
\usepackage{a4wide}

\def\squareforqed{\hbox{\rlap{$\sqcap$}$\sqcup$}}
\def\qed{\ifmmode\squareforqed\else{\unskip\nobreak\hfil\penalty50\hskip1em\null\nobreak\hfil\squareforqed\parfillskip=0pt\finalhyphendemerits=0\endgraf}\fi}

\newtheorem{definition}{Definition}[section]
\newtheorem{theorem}{Theorem}[section]
\newtheorem{lemma}{Lemma}[section]
\newtheorem{corollary}{Corollary}[section]
\newtheorem{claim}{Claim}
\newtheorem{subclaim}{Subclaim}

\newtheorem{proof}{Proof}

\newcommand{\comment}[1]{}
\newcommand{\N}{\mathbb{N}}
\newcommand{\0}{\emptyset}
\newcommand{\ind}{\mathrm{ind}}
\newcommand{\harp}{\upharpoonright}
\newcommand{\pl}{\text{PL}}
\newcommand{\pdeep}{\text{-deep}_{\text{PL}}}
\newcommand{\Pdeep}{\text{deep}_{\text{PL}}}
\newcommand{\cp}{\text{P}}
\newcommand{\dt}{\text{DTIME}}
\newcommand{\cnp}{\text{NP}}
\newcommand{\cam}{\text{AM}}
\newcommand{\cexp}{\mathrm{EXP}}
\newcommand{\ce}{\mathrm{E}}
\newcommand{\low}[2]{\mathrm{Low({#1},{#2})}}
\newcommand{\high}[2]{\mathrm{High({#1},{#2})}}

\begin{document}

\title{Polylog Depth, Highness and Lowness for $\ce$}

    \author{Philippe Moser
    \footnote{Computer Science Department, National University of Ireland Maynooth, Maynooth, co. Kildare, Ireland}}

\maketitle
\begin{abstract}
    We study the relations between the notions of highness, lowness and logical depth in the setting of complexity theory. We introduce a new notion of polylog depth based on time bounded Kolmogorov complexity. We show polylog depth satisfies all basic logical depth properties, namely sets in $\cp$ are not polylog deep, sets with (time bounded)-Kolmogorov complexity greater than polylog are not polylog deep, and only polylog deep sets can polynomially Turing compute a polylog deep set. We prove that if  $\cnp$ does not have $p$-measure zero, then $\cnp$ contains polylog deep sets. We show that every high set for $\ce$ contains a polylog deep set in its polynomial Turing degree, and that there exist $\low{\ce}{\cexp}$ polylog deep sets. Keywords: algorithmic information theory; Kolmogorov complexity; Bennett logical depth.
\end{abstract}

\section{Introduction}

Bennett \cite{b:bennett88} introduced the notion of logical depth, as an attempt to characterise \emph{useful} information (e.g. the halting problem) as opposed to non-useful information (e.g. a Martin-L\"of random sequence).
Interestingly, non-deep sequences (called shallow sequences) can range from trivially organised sequences such as 01010101..., to completely unorganised ones such as a random sequence. Logically deep sequences are somewhere between these two extremes.

 A sequence is Bennett deep \cite{b:bennett88} if the difference of every computable approximation of the Kolmogorov complexity of its initial segments, and the actual Kolmogorov complexity, is unbounded. This difference is called the depth magnitude of the sequence \cite{DBLP:journals/dmtcs/MoserS17}.

Moser and Stephan \cite{DBLP:journals/dmtcs/MoserS17} studied the differences in computational power of sequences of different depth magnitudes, within the context of computability theory. They related logical depth to standard computability notions (e.g. highness, diagonally-non-computability and lowness). 	Highness and lowness are important characterisations of the computational power of  sets used in computability theory \cite{odifreddi}. Informally a set is high (resp. low) if it is useful (resp. not useful) given as an oracle.
	Among others, they showed that a Turing degree is high iff it contains a deep set of large depth magnitude \cite{DBLP:journals/dmtcs/MoserS17}. They found that not all deep sets need be high, by constructing a low deep set 
(subsequently refined by Downey, McInerney and Ng into a low r.e. deep set \cite{DBLP:journals/tcs/DowneyMN17}).

In this paper, we revisit the results of Moser and Stephan \cite{DBLP:journals/dmtcs/MoserS17} within the context of computational complexity theory. 
There are two main difficulties in trying to do so. Firstly
the quest for the ``right'' polynomial version of depth has turned out to be elusive. There are several issues inherent to complexity theory that are hard to overcome, while trying to translate logical depth from computability to complexity theory. It is non-trivial to find a polynomial depth notion that simultaneously satisfies all natural properties of Bennett's original notion. Secondly,  translating the results from \cite{DBLP:journals/dmtcs/MoserS17} to complexity theory is not straightforward, because the main techniques in \cite{DBLP:journals/dmtcs/MoserS17} do not scale down to complexity theory. Several computability techniques used in  \cite{DBLP:journals/dmtcs/MoserS17}
(e.g. the equivalence between highness and dominating functions, the notion of DNC degrees) have no natural equivalent concept in complexity theory.
 Therefore the proof of the main result in this paper (the existence of low deep sets) rely on different techniques from the ones used in \cite{DBLP:journals/dmtcs/MoserS17}.

As mentioned above, finding a good polynomial depth notion is not straightforward. There have been several attempts by various authors over the last decade  \cite{b.antunes.depth.journal,DBLP:conf/cie/DotyM07,DBLP:journals/tcs/Moser13}.
The main difficulty is to achieve a notion that  satisfies the natural properties of Bennett's logical depth (among others: easy and hard sets are not deep, non-deep sets can't compute deep ones, some deep set exists), while at the same time is powerful enough to prove interesting results. All notions so far have been a compromise between these two goals, i.e. relaxing the satisfaction of some of the natural properties in order for the notion to be useful. 
See   \cite{DBLP:journals/tcs/Moser13} for a summary of previous polynomial depth notions, and the compromises made by each notion.

In this paper we study a polylog version of depth as close as possible to the original notion by Bennett, namely the difference of two Kolmogorov complexities with different time bounds. Informally Bennett's depth measures some aspect of the difference in power between $\Delta^0_2$ and $\Delta^0_1$. In our polylog setting, this becomes $\cexp$ vs $\cp$. This corresponds to quasipolynomial  vs polylog in the setting of Kolmogorov complexity, because the size of the characteristic sequence is exponentially larger than the size of the strings it encodes (i.e. 
sets in $\cp$ have their polynomial time complexity measured relative to the  size of the input, which corresponds to polylog with respect to the size of the characteristic sequence). To allow for Kolmogorov complexity with polylog time bounds, i.e. such that there is not enough time to read the whole program, we use the oracle Kolmogorov complexity model of \cite{DBLP:journals/siamcomp/AllenderBKMR06}. This model is equivalent to standard Kolmogorov complexity for time bounds linear or greater, but  allows for  sublinear time bounds. 

Polylog depth is interesting because it interacts nicely with lowness and highness for the complexity class $\ce$, as explained below. But similarly to previous polynomial depth notions, this comes at a cost (the compromise):   the natural property ``random implies non-deep'' is not formulated in terms of the canonical random sets for $\cexp$ (the ones on which no $p_2$-martingale succeeds), instead we use ``complex'' sets (sets for which the  time bounded Kolmogorov complexity is large). It is currently unknown whether  $p_2$-random sets and ``complex'' sets are the same, see \cite{DBLP:journals/jcss/HitchcockV06} for more on this question. 
Apart from this compromise, polylog depth satisfies the natural properties of depth,
namely ``easy'' sets (here: in $\cp$)  and ``random'' enough sequences (here: complex sets) are shallow,
and polylog depth satisfies a  slow growth law, i.e. no shallow sequence can quickly compute a deep one. 
Also $\cexp$ contains polylog deep sets. 

Next we investigate whether $\cnp$ contains any polylog deep sets. Since one cannot exclude $\cp=\cnp$ (in which case all $\cnp$-sets are shallow), one can only hope for a conditional result. We prove that if $\cnp$ does not have $p$-measure zero, then it contains a polylog deep set. The assumption $\cnp$ does not have $p$-measure zero is a reasonable assumption based on Lutz's $p$-measure \cite{Lutz:almost_everyw_high_nonunif_complex}, which has implications not known to follow from $\cp\neq\cnp$. Examples include separating many-one and Turing reductions \cite{DBLP:journals/tcs/LutzM96} and derandomization of $\cam$ \cite{DBLP:conf/fsttcs/ArvindK97,DBLP:journals/iandc/ImpagliazzoM09}, see \cite{lutz-quantitative-structure-of-exp} for more. 

Next we study the relation between highness and depth, and show that each set that is high for $\ce$ (i.e. a set $A$ such that $\ce^\ce\subseteq\ce^A$) contains a polylog deep set in its polynomial Turing degree. This is on par with the results of \cite{DBLP:journals/dmtcs/MoserS17} at the polylog level.
	 The idea of the proof is that highness enables the set to compute polynomially random strings of small sizes, but large enough to guarantee depth of the whole sequence.
	
Our main result investigates whether all polylog deep sets need be high. We find a negative answer, by constructing a polylog deep set in $\low{\ce}{\cexp}$, (i.e. sets $A$ such that $\ce^A\subseteq \cexp$). Lowness for $\ce$ (i.e. $\low{\ce}{\ce}$)  was first studied in \cite{DBLP:journals/siamcomp/BookORW88}, where a low set in $\ce -\cp$ was constructed. The set constructed in \cite{DBLP:journals/siamcomp/BookORW88} is a very sparse set of
	strings that are random at the polynomial time level but not at the exponential time level. The sparseness of the set guarantees that large queries can be answered with ``no'', hence only small queries need to be computed, which guarantees lowness. However the set in \cite{DBLP:journals/siamcomp/BookORW88} is too sparse to be polylog deep. 
We construct a new set   with blocks of subexponential size each containing a random string. The size of the blocks are measured by a tower of subexponential functions,
in order to be able to satisfy two conflicting requirements, namely the sequence needs to be sparse enough to stay low, but the relative size of blocks needs to  be small enough so
that the polylog depth of each block is preserved over the whole sequence.

\section{Preliminaries}
	Logarithms are taken in base $2$ and rounded down. For simplicity of notation we write $\log n$ for $\lfloor\log_2 n\rfloor$ and 
	$\log^{(2)}n$ stands for $\log\log n$. By convention, whenever a real number is cast to an integer, we take the floor of the real number; we omit the floor notation for simplicity of notation.
	
	We use standard complexity/computability/algorithmic randomness theory notations 
	(see \cite{Diaz:structural_complexity1,Diaz:structural_complexity2,odifreddi,downey:book,nies:book}).
	We write $\{0,1\}^n $ for the set  of strings of size $n$.
	We denote by $s_0,s_1,\ldots,s_n$ the standard enumeration of strings in lexicographic order.
	For a string $x$, its length is denoted by $|x|$. The empty string is $s_0=\epsilon$.
	The index of string $x$ is the integer $\ind(x)$ such that $x=s_{\ind(x)}$. For every natural number $n$, it holds $|s_n| = \log (n+1) $, and the index of strings of length $n$ are in the interval $[2^n-1,2^{n+1}-2]$. We identify $n$ with $s_n$, in particular $|n|=|s_n|=\log (n+1)$.
	We say string $y$ is a prefix of string $x$, denoted $y\prec x$, if there exists a string $a$ such that
	$x=ya$. 

	A sequence is an infinite binary string, i.e. an element of $2^{\omega}$.
	For string or  sequence  $S$  and $i,j\in  \mathbb{N}$, we write $S[i, j]$  for the string consisting of
	the $i^{\textrm{th}}$ through $j^{\textrm{th}}$  bits of $S$, with
	the conventions that $S[i, j]=\epsilon $ if $i>j$, $S[i]=S[i,i]$, and $S[0]$ is the
	leftmost bit of $S$. We write $S\harp i$ for $S[0, i-1]$ (the first $i$ bits of $S$) and $S\harp x$ for $S\harp \ind(x)$.
	The characteristic sequence of a set of strings $L$ is the sequence $\chi_L \in 2^{\omega}$,
	whose $n$th bit is one iff $s_n \in L$. We abuse the notation and use  $L$ and  $\chi_L$ interchangeably.
	Note that for any string $x$, $|L\harp x|=O(2^{|x|})$.

		We consider the following standard time bound families: $\cp= \{  kn^k| \ k\in\N  \}$, 
	$\pl= \{ k \log^k n|\ k\in\N\}$, $\ce = \{  2^{kn}  |\ k\in\N \}$, and $\cexp = \{  2^{n^k}|\ k\in\N\}$.
	We abuse notations by using time bound families for complexity classes interchangeably, e.g. $\ce = \cup_{c\in\N}\dt(2^{cn})$.
	Logarithmic time bounds require oracle access to the input as explained below.

	We use $\leq^+$ (resp. $=^+$) to denote less or equal (resp. equal) up to a constant term. We fix
	a poly-computable 1-1
	pairing function $\langle \cdot \rangle : \N\times\N\rightarrow\N$.
	An order function is an unbounded non-decreasing function from
	$\mathbb N$ to $\mathbb N$, computable in polynomial time, e.g. $\log n , n\log n, n^2$.

	We consider standard polynomial Turing reductions $\leq_T^p$.
	Two sets $A,B$ are polynomial Turing equivalent ($A\equiv_T^p B$) if $A\leq_T^p B$
	and $B\leq_T^p A$.
	The polynomial Turing degree of a set $A$ is the class of sets polynomial Turing equivalent to $A$.

	Fix a universal  Turing machine $U$.
	The (plain) Kolmogorov complexity of string $x$, denoted $C_U(x)$,
	is the length of the lexicographically first
	program $x^*$ such that $U$  on input $x^*$ outputs $x$. It can be
	shown that the value of $C_U(x)$  does not depend on the choice of 
	$U$  up to an additive constant, therefore we drop $U$ from the notation and write $C(x)$. 
	$C(x,y)$ is the length of a shortest program that outputs the pair $\langle x,y\rangle$,
	and $C(x\ |\ y)$ is the length of a shortest program such that $U$ outputs $x$
	when given $y$ as an advice. 

	We also consider time bounded Kolmogorov complexity. To allow logarithmic time bounds i.e. shorter than the time required to read the full program, we use the oracle model of \cite{DBLP:journals/siamcomp/AllenderBKMR06}. In this model, the universal machine is provided with program $p$ as an oracle (written $U^p$), and 
	can query any bit of it.
	As noticed in \cite{DBLP:journals/siamcomp/AllenderBKMR06}, the definition coincides with the standard time bounded Kolmogorov complexity, for time bounds greater than $O(n)$.
	Given time bound $t\geq\log n$, define 
	$$C^t(x)= \min\{ |p|:\ \forall b\in \{0,1,\epsilon\} \ \forall i \leq n, U^p(i,b)[t(n)]\downarrow = \text{accepts iff } x[i]=b\} $$ 
	where $n=|x|$, $x[0], x[1],\ldots, x[n-1]$ are the $n$ bits of $x$, $x[m]=\epsilon$  for all $m\geq n$ and $U^p(i,b)[t(n)]\downarrow$ means $U$ with oracle access to $p$ halts within $t(n)$ steps on input $i,b$.
	
	To concatenate two programs $p,q$ one can double all the bits in $p$, and append $01$. We write $\hat p$ for this string, i.e. from $\hat p q$ it is easy to recover $p,q$.

	A sequence $A$ is called $(\pl,\pl)$-complex if for all $k,n \geq 0$, $C^{\log^k n} (A \harp n) >^+ \log^k n$. We use the same $(\cdot, \cdot)$ notation for other levels of complex sets, where the first (resp. second) parameter is the time bound (resp. the minimal program size bound).

		The symmetry of information holds for exponential time bounds.

	\begin{theorem}[Symmetry of information \cite{b.vitanyi}, 7.1.12 p.548]\label{t.sym.inf}
	Let $t\in\ce$ be a time bound. Then there exists time bound $t'\in\ce$ such that for any strings $x,y$, we have $C^t(x,y)\geq C^{t'}(x) + C^{t'}(y|x) - O(\log(|x|+|y|))$.
	\end{theorem}

	 Lutz \cite{Lutz:q_structure_of_EXP} used Lebesgue measure to define a measure notion on complexity classes e.g. $\ce,\cexp$.
\comment{
	A martingale is a function $d:\{ 0,1 \}^{*} \rightarrow [0,\infty[$ such that,
	$$d(w) = \frac{d(w0) + d(w1)}{2} $$
	for every $w \in \{ 0,1 \}^{*}$.
	We say $d$ is a $p$-martingale (sometimes written $p_1$-martingale) if $d$ is computable in time polynomial in $|w|$, and
	$d$ is a $p_2$-martingale if $d$ is computable in time $2^{\pl(|w|)}$.

	This definition corresponds to  the following  betting game in which a gambler
	puts bets on the successive membership bits of a hidden language $A$.
	At the end of round $n$,
	it is revealed to the gambler whether $s_{n} \in A$ or not.
	The gambler starts with capital 1.
	Then, in round $n$, depending on the first $n-1$ outcomes
	$w=A\harp n$, the gambler bets a certain fraction
	$\epsilon_{w}d(w)$ of his current capital $d(w)$, that the $n$th word
	$s_{n} \in A$, and bets the remaining capital $(1-\epsilon_{w})d(w)$ on
	the complementary event $s_{n} \not \in A$.
	The game is fair, i.e. the amount put on the correct event is doubled,
	the one put on the wrong guess  is lost.
	The value of $d(w)$, where $w = A\harp n$ equals the
	capital of the gambler after round $n$ on language $A$. The  play proceeds forever. 

	The player wins on a language $A$ if he manages to make his capital
	arbitrarily large during the game.
	We say that a martingale $d$ succeeds on a language $A$, if
	$d(A) := \limsup_{n\rightarrow\infty} d(A\harp n) = \infty$.
	The success set
	$S^{\infty}(d)$ of a martingale $d$ is the class of all languages on which $d$ succeeds.
}

	For the rest of this section, let $i\in\{1,2\}$; $p_1$ (abbreviated $p$) stands for polynomial time bounds, $p_2$ for $2^{\pl}$.
	\begin{definition}[See \cite{Lutz:q_structure_of_EXP} for more details] 
		A class $C$ has $p_i$-measure zero (written $\mu_{p_i}(C)=0$) if there is a single $p_i$-martingale $d$
		that succeeds on every language $A$  of $C$.
	\end{definition}
	This property is monotone in the following sense:
	If class $D$ is contained in a class $C$
	of $p_i$-measure zero, then $D$ also has $p_i$-measure zero. 
	It is easy to see that if a class $C$ has $p_1$-measure zero, then it has $p_2$-measure zero. The converse is not always true but the following  is known.

	\begin{theorem}[Juedes and Lutz \cite{DBLP:journals/tcs/JuedesL95}\label{l.property2}]
	NP has $p_2$-measure zero iff NP has $p$-measure zero.	
	\end{theorem}

	The measure conservation property \cite{Lutz:almost_everyw_high_nonunif_complex} states that  class $\ce$ (resp. $\cexp$) does not have $p$-measure (resp. $p_2$-measure) zero.
\comment{
	Uniform unions of measure zero sets have measure zero, as the following results shows
	\begin{lemma}[Lutz \cite{Lutz:almost_everyw_high_nonunif_complex}] \label{t.property1}
		Suppose $\{ d_{j} \}_{j \geq 1}$ is a set of $p$-martingales (resp. $p_2$-martingales), where $d_j$ is computable in time $(n+j)^k$ (resp. $2^{\log^k (n+j)}$) (for some $k\in\N$), and $\sum_i d_i(\epsilon)<\infty$.	
	Then there exists a martingale $d$ computable in time $n^c$ (resp. $2^{\log^c n}$), for some $c\in\N$, such that for all $i$ and all strings $w$, $d(w)\geq 2^{-i}d_i(w)$ and $d(\epsilon)<\infty$.
	\end{lemma}
}

	Lutz \cite{Lutz:almost_everyw_high_nonunif_complex} showed that sets with small circuit complexity have $p_2$-measure zero, which is implied by the following result.
	\begin{theorem}[Lutz \cite{Lutz:almost_everyw_high_nonunif_complex}] \label{t.property1}
	Let $c\in\N$. The set $\{A\in 2^{\omega}|\ \exists^{\infty} n \ C^n(A\harp n) < \log^c n\}$ has $p_2$-measure zero.
	\end{theorem}

\section{Polylog depth}
	A sequence $S$ is  $h\pdeep$ if the difference between the large and small time-bounded Kolmogorov complexity of the prefixes of $S$ exceeds some order function $h$. We call the order $h$ the depth magnitude of $S$. As shown in \cite{DBLP:journals/dmtcs/MoserS17}, the choice of $h$ can have consequences on the computational complexity of $S$. 
	\begin{definition}\label{d.d1}
	Let $m(n) \leq n$ be an order function. A set $S$ is $m\pdeep$ if 
	for every $t\in\pl$ there exists $t'\in 2^{\pl}$ such that for almost every $n$,  $C^t(S\harp n) - C^{t'}(S\harp
	n) \geq m(n)$.
	\end{definition}
	We say $S$ is $O(1)\pdeep$ if it is $c\pdeep$ for every $c\geq 0$.
	We say $S$ is $\pl\pdeep$ if it is $m\pdeep$ for every $m \in \pl$ (and similarly for other order function families).

	In the following section, we show that polylog depth satisfies most natural properties (scaled down to complexity theory) of Bennett's original notion.
	Firstly, as noticed in \cite{DBLP:journals/tcs/Moser13}, for most depth notions it can be shown that easy and random sequences are not deep. The following two results show that this is also the case for $\Pdeep$.

    \begin{theorem}\label{t.pnotdeep}
	Let  $A$ be in $\cp$. Then $A$ is not $2\log n\pdeep$. 
	\end{theorem}
	We need the following lemma; it follows the same proof as the unbounded Kolmogorov complexity case, but since we are considering sublinear time bounds, we need to check it  still holds.
	\begin{lemma}\label{l.1}
	Let $A\in\dt(n^c)$ (resp. $\dt(2^{n^c})$). For every $t\in\pl$ (resp. $t\in2^{\pl}$) and $n\in\N$, $C^{t'}(A\harp n) \leq^+ C^t(1^n)$ with $t'(n)=O(t(n)+\log^c n)$ (resp. $t'(n)=O(t(n)+2^{\log^c n})$.
	\end{lemma}
    \textbf{Proof} (of Lemma \ref{l.1}).
	Let $A,t$ be as above and let $n\in\N$ and $M$ be a TM deciding  $A$ in time $n^c$.
	Let $p$ denote a minimal $t$-program for $1^n$, i.e. $|p|=C^t(1^n)$. Consider prefix-free program $p'=\hat q p$, where $q$ are instructions  such that for $b= 0,1,\epsilon$
	$U^{p'}(i,b)$ simulates $U^p(i,1)$. If $U^p(i,1)$ accepts (i.e. $i<n$), simulate $M(s_i)$ and accept iff $M(s_i)=b$. If $U^p(i,1)$ rejects, then accept iff $b=\epsilon$. The first step takes at most  $t(n)$ steps, and the simulation of $M$ on an input of size at most $\log n$ takes at most  $\log^c n$ steps, for a total of $O(t(n)+\log^c n)$ steps. The exponential case is similar.
	\qed

   ~\newline
    \textbf{Proof} (of Theorem \ref{t.pnotdeep}).
	Let us prove the theorem. Let $A\in\cp$, decidable in time $n^k$, let $t(n)=O(\log n + \log^k n)$ be given by Lemma \ref{l.1}, and let $t'\in 2^{\pl}$.  For every $n\in\N$,
	\begin{align*}
	C^t(A\harp n) - C^{t'} (A\harp n)  &\leq^+ C^{2\log n}(1^n) \qquad & \text{By Lemma \ref{l.1}}\\
	&\leq^+ \log n &
	\end{align*}
      where the last inequality holds by a print program (i.e. a program $p = \hat q e$ where $q$ are instructions, $e$ is a binary encoding of $n$, 
        such that for $b\in \{ 0,1,\epsilon\}$,
      $U^{p'}(i,b)$ accepts iff ($i\leq n$ and $b=1$) or  ($i > n$ and $b=\epsilon$).
      Therefore $A$ is not $2\log n\pdeep$, which ends the proof.
	\qed
~\\
	From the point of view of  Bennett's logical depth, shallow sequences are those whose structure is either very organised, e.g. the sequence 1010101010..., or lack any organisation at all, e.g. a random sequence. For the latter case, Bennett \cite{b:bennett88} showed that
	Martin-L\"of random sequences are not deep.  A similar result holds in our setting: 
    \begin{lemma}
	Let $A$ be $(2^{\pl}, n - 2\log n)$-complex, then $A$ is not $2\log n\pdeep$. 
	\end{lemma}
	\begin{proof}
	Let $A$ be as above i.e.,  for every $k\in\N$, for almost every $n$ 
	\begin{equation}\label{e.1}
	C^{2^{\log^k n}}(A\harp n)>n-2\log n.
	\end{equation}
	Since $C^{2\log n} (A \harp n) \leq^+ n$ via a ``print'' program, we have 
	$C^{2\log n} (A \harp n) - C^{2^{\log^k n}}(A\harp n)\leq^+ n - (n -  2\log n)$,
	i.e. $A$ is not $2\log n \pdeep$.
	\qed
	\end{proof}
~\\
Note such sequences exist:  every Martin-L\"of sequence satisfies $C(A\harp n) \geq n - K(n) - O(1) > n - 2 \log n$; and not all  such sequences are Martin-L\"of (e.g. see \cite{downey:book} section 8.4: a no gap theorem for 1-randomness, for more details).

	Bennett showed that producing a logical deep sequence requires a complex and lengthy computation \cite{b:bennett88}.
	For example truth-table reductions are not capable of computing a Bennett deep sequence from the empty set. This is known as the slow growth law,
	which states that  if some set $A$ truth-table computes some deep set $B$, then $A$ is deep.
	A similar result holds for $\Pdeep$,  once the power of the corresponding reductions is adapted to the polynomial world. Truth-table reductions are 
	too powerful in the setting of complexity theory (they can compute every set in every complexity class included in $\Delta^0_1$), thus they need to be replaced with
	polynomial time Turing reductions.

	\begin{theorem}\label{t.sgl}
	Let  $A,B\in\cexp$, $A\leq_T^p B$ in time $O(n^c)$, and $A$ is $\alpha\log^{ck}n\pdeep$ for some $k  , \alpha>0$, then $B$ is $\beta\log^{k}n\pdeep$ for some $\beta>0$.
	\end{theorem}
	\begin{lemma}\label{l.66}
	Let $A\leq^p_T B$, where $en^c$ is the running time of the reduction. For every $s'\in\pl$ (with $s'=\log^k n$) and $n\in\N$ we have
	$$
	C^s(A\harp 2^{a\log^{1/c}n })\leq^+ C^{s'}(B\harp n) + 2\log n
	$$
      for some $a >0$ and  $s(n)= \frac{2}{a^{ck+c}} \log^{ck+c}n$.
	\end{lemma}
    \textbf{Proof} (of Lemma \ref{l.66}).
	Let $A,B,c,e,s'$ be as above, $M$ be the TM computing the reduction,  and $n\in\N$.
	Let $p$ be a minimal $s'$-program for $B\harp n$. Consider the following program $p'$ for $A\harp 2^{a\log^{1/c}n }$, where
        $p'=\hat q \hat q' p $, $q$ is a set of instructions, $q'$ encodes $n$ (i.e. $|\hat q'|\leq^+ 2\log n)$, and where
	$U^{p'}(i,b)$ does the following: 
        \begin{itemize} 
            \item Recover $n$. 
            \item If $i\geq  2^{a\log^{1/c}n}$, accept iff $b=\epsilon$. 
            \item If $i<  2^{a\log^{1/c}n}$,  simulate $M^B(s_i)$, answering each query $s_j$ to $B$ by simulating $U^p(j,0)$. 
                Accept iff $b= M^B(s_i)$.
        \end{itemize}
                For the last item above, notice that all queries are within $B\harp n$,
	because $A\harp 2^{a\log^{1/c}n}$ codes for strings of length at most $a\log^{1/c}n$. Therefore the largest query has size at most  $e(a\log^{1/c}n)^c = ea^c \log n$, 
	i.e. a string with index at most $2^{1+ea^c\log n}-2< 2^{2ea^c\log n} < n$, by choosing $a$ small enough.
	The simulation of $M^B(s_i)$ takes at most $e|s_i|^c\leq\log n$ steps, and there are at most $\log n$ queries to $B\harp n$, each requiring at most $s'(n)$ steps, thus a total of $s'(n)\log n+\log n\leq 2s'(n)\log n$. 
        Let us express this time bound as a function of the input size $m=2^{a\log^{1/c}n}$. We have $\frac{1}{a^c}\log^{c}m=\log n$, i.e. $2^{\frac{1}{a^c}\log^{c}m}=n$ hence the total number of steps becomes
      $$2s'(2^{\frac{1}{a^c}\log^{c} m})\frac{1}{a^c}\log^{c}m =\frac{2}{a^{ck+c}}\log^{c+ck} m.$$
	\qed
~\\

    \textbf{Proof} (of Theorem \ref{t.sgl}).
	Let us prove the theorem. Let $A,B,\alpha,k$ be as above, and $en^c$ be the running time of the $\leq^p_T$-reduction,   and $\beta >0$ to be determined later.
	By contradiction, suppose $B$ is not $\beta\log^{k}n\pdeep$ i.e., there exists $t=\log^d n$ such that for every $t'\in 2^{\pl}$ there exists an infinite set $N$ such that for every $n\in N$, $C^t(B\harp n) < \beta\log^{k} n + C^{t'}(B\harp n)$.
        Suppose $B$ is decidable in time $2^{n^b}$ (for some $b>0$). Consider $\bar t(n)=\frac{2}{a^{cd+c}} \log^{cd +c}n$, where $a$ is given by Lemma \ref{l.66}, and $t'(n)=n+2^{\log^b n}$, and let $n\in N$ where $N$ is  the infinite set of lengths testifying the non depth of $B$ for this $t'$. We have
	\begin{align*}
      C^{\bar t}(A\harp 2^{a\log^{1/c}n}) &\leq^+ C^t(B\harp n)  +2\log n \qquad &\text{By Lemma \ref{l.66}}\\
      &\leq \beta\log^{k} n + C^{t'}(B\harp n) +2\log n \qquad &\text{Because }n\in N \\
      &\leq^+ \beta\log^{k} n +C^n(1^n) +2\log n \qquad &\text{By Lemma \ref{l.1}}\\
	&\leq^+ \beta\log^{k} n +\log n +2\log n \\
        &< 4\beta\log^{k}n.
	\end{align*}
	Thus for all $n\in N$ and all $s\in 2^{\pl}$,
	\begin{align*}
        C^{\bar t}(A\harp 2^{a\log^{1/c}n}) - C^{s}(A\harp 2^{a\log^{1/c}n})&\leq  C^{\bar t}(A\harp 2^{a\log^{1/c}n}) -0  < 4\beta\log^{k}n = \frac{4\beta}{a^{ck}} \log^{ck}(2^{a\log^{1/c}n}) 
	\end{align*}
        i.e. $A$ is not $\frac{4\beta}{a^{ck}}\log^{ck}n\pdeep$. Choosing $\beta>0$ small enough such that $4\beta/a^{ck}< \alpha$, we get a  contradiction.
	\qed
~\\

	The proof above implicitly shows the following: $(\pl, \pl)$-complex sets are closed upwards under poly Turing reductions. All $\pl\pdeep$ sequences are  $(\pl, \pl)$-complex. The converse holds if the sequence is in $\cexp$. It would be interesting to see whether Theorem \ref{t.sgl} still holds without the assumption that the set is in $\cexp$, though this is currently unknown.

	By enumerating all short $O(n)$-time programs, one can for every $k>1$  construct a $(\pl,\log^{k+1}n)$-complex sequence in EXP, i.e. a sequence $A\in\cexp$ such that 
	for all   $n\in\N$, $C^n(A\harp n)>\log^{k+1} n$. Since $A\in\cexp$, by Lemma \ref{l.1}, $A$ is $\log^{k} n\pdeep$. 
\comment{currently false: trying to use log^2 n complex set with n^2 poly reduction with k = 1 in the slow growth law: there's not enough "room" to get a deep set. increasing the complexity forces to increase the reducibility time and we gain nothing.
	Bennett showed in \cite{b:bennett88} that the halting problem is deep. Similarly, we show that the canonical EXP-complete set is deep.
	The proof is a consequence of the slow growth law.
	\begin{corollary}\label{c.exp.complete}
	$H_{\cexp}$ is  $\beta\log n\pdeep$ for some $\beta>0$.
	\end{corollary}
	\begin{proof}
	In EXP, one can construct a set $R$ with $C^n(R\harp n > \log^2 n$ by simulating all $2^{\log^2 n}$ short programs for $n$ steps. Thus $R\in\dt(2^{|x|^2})$, i.e. $R$ is reducible to $H_{\cexp}$ in time $en^2$ for some $e>0$.
	Thus $R$ is $1/2\log n\pdeep$ because for every $t\in\pl$ there exists $t'\in 2^{\pl}$ (given by Lemma \ref{l.1}) such that $C^t(R\harp n) - C^{t'}(R\harp n) \geq^+ \log^2 n - \log n > 1/2 \log n$.
	By Theorem \ref{t.sgl} with $k = 1, c = 2, \alpha = 1/2$, $B$ is $\beta\log n\pdeep$ for some $\beta>0$
	\qed
	\end{proof}
}
~\\

	It is natural to ask whether some NP sets are $\Pdeep$. One cannot exclude the possibility that $\cp = \text{NP}$, in which case the answer is negative,
	but if one assumes that NP is not a small subset of EXP, then one can show that NP contains deep sets.
	To measure the size of NP within EXP, we use Lutz's $p$-measure \cite{Lutz:almost_everyw_high_nonunif_complex}, which is 
	a complexity version of Lebesgue measure that allows to measure the size of subsets of EXP. Thus the statement that  NP is a not a small subset of EXP, is formalised by the statement that  NP does not have $p$-measure zero. The assumption that NP does not have $p$-measure zero implies P$\neq$NP and is not known to follow from it. It has been used to show some interesting results, e.g.\cite{DBLP:journals/tcs/LutzM96,DBLP:conf/fsttcs/ArvindK97,DBLP:journals/iandc/ImpagliazzoM09,lutz-quantitative-structure-of-exp}, that are not known to follow from P$\neq$NP.
	\begin{theorem}\label{t.nonzero}
        If $C\subseteq\cexp$ with $\mu_{p_2} (C)\neq 0$ then $C$ contains a $\log^k n\pdeep$ set for every $k>0$.
	\end{theorem}
	\begin{proof}
        Suppose $\mu_{p_2} (C)\neq 0$ and $C$ does not contain a $\log^k n\pdeep$ set for some $k>0$.
	Then for every $A\in C$  and for every $t'\in 2^{\pl}$ there exists an infinite set $N$, such that for every $n\in N$,
	$C^t(A\harp n)- C^{t'}(A\harp n)\leq  \log^k n$. Thus
	$C^t(A\harp n)\leq \log^{k+1} n$, by Lemma \ref{l.1}, since $A\in\cexp$. Thus for every $A\in C$, there exists an infinite set $N_A$, such that for every $n\in N_A$, $C^n(A\harp n)<\log^{k+1} n$.
	Thus by Theorem \ref{t.property1}, $\mu_{p_2}(C)=0$; a contradiction. 		\qed
	\end{proof}
    ~\\
\begin{corollary}
	If $\mu_p (\cnp)\neq 0$ then NP contains a $\log^k n\pdeep$ set for every $k>0$.
\end{corollary}
\begin{proof}
    Apply Theorem \ref{t.nonzero} together with Theorem \ref{l.property2}.
    \qed
\end{proof}

\section{Highness and depth}

	Highness and lowness are important characterisations of the computational power of  sets used in computability theory \cite{odifreddi}. Informally a set is high (resp. low) if it is useful (resp. not useful) given as an oracle.
	The notions were generalised to considering two classes, with applications in randomness theory \cite{downey:book,nies:book}.
	Complexity versions of these notions were developed for NP \cite{Diaz:structural_complexity1,Diaz:structural_complexity2}, and E \cite{DBLP:journals/siamcomp/BookORW88}.
	\begin{definition}
		Let $C\subseteq D$ be two complexity classes.
		\begin{enumerate}
		\item Set $A$ is $\low{C}{D}$ if $C^A\subseteq D$.
		\item Set $A$ is $\high{C}{D}$ if $C^A\supseteq D^D$.
		\end{enumerate}
		Set $A$ is low for $\ce$ if it is $\low{\ce}{\ce}$.
		Set $A$ is high for $\ce$ if it is  $\high{\ce}{\ce}$.
	\end{definition}

	Lowness and highness for E are preserved under polynomial Turing reductions.

	\begin{lemma}\label{l.6}
	Let $A$ be high (resp. low) for $\ce$. Then all sets in the polynomial Turing degree of $A$ are high (resp. low) for $\ce$.
	\end{lemma}
	\begin{proof}
	Let $A$ be high for $\ce$, $B$ be in the $\le^p_T$-degree of $A$ and $L\in\ce^{\ce}$. By highness for $\ce$ of $A$, there is an oracle TM $M^A$ that decides $L$ in time $2^{an}$. On an input $x$ of size $n$,
	$M$ makes at most $2^{an}$ queries to $A$, each of size less than $2^{an}$. Since $A,B$ are in the same $\leq^p_T$-degree, there is a machine $N^B$ deciding $A$ in time $n^b$. Thus each query $q$ to $A$ can be answered  by $N^B$
	in $|q|^b \leq (2^{an})^b = 2^{abn}$ steps. Thus $L$ can be decided in time $2^{abn}2^{an}\leq 2^{2abn}$ with oracle access to $B$, i.e. $L\in\ce^B$.
	\noindent
	The proof for lowness is similar.
	\qed
	\end{proof}
	
    ~\\

	The following result shows that any high set computes a deep set. The idea of the proof is that highness enables the set to compute polynomially random strings of small sizes, but large enough to guarantee depth of the whole sequence.
	\begin{theorem}\label{t.high}
	Let $A$ be high for $\ce$. Then   the $\leq^p_T$-degree of $A$ contains a set that is $\log^{(2)}n\pdeep$.
	\end{theorem}
	\begin{proof}
	Let  $c\in\N$ to be determined later. Consider the following set of random strings $R=\{ x|\ C^{2^{2^{3n}}}(x) \geq |x|\}\in \ce^{\ce}\subseteq\ce^A$. Let $j\in\N$, $j'=2^{1+ \log j^c }$ so that $j^c\leq j'\leq2j^c$ and $\log j' = 1+ \log j^c $.
	Define
	$$B[2^j-1,2^{j+1}-2]=R[j'-1,2j'-2]A[2^{j-1}-1,2^j-2]0^{2^j-j'-2^{j-1}}$$
        i.e. $B\cap \{0,1\}^j$ codes for $R\cap \{0,1\}^{\log j'}$ and $A\cap \{0,1\}^{j-1}$.
	\begin{claim}
	$B$ is in the $\leq^p_T$-degree of $A$.
	\end{claim}
	For each $x$ of length $\log j'$, deciding whether $x\in R$ requires at most $2^{a|x|}$ queries of size at most $2^{a|x|}$ to $A$ (for some $a\in\N$), i.e. $2^{a\log j'}\leq O(j^{ac})$ queries of size 
	$2^{a\log j'}=O(j^{ac})$. Since there are $j'$ such $x$'s, we have  $R[j',2j'-2]$ can be computed in at most $O(j^{2ca})$ steps,  hence $B\leq^p_T A$. Since $A\leq^p_T B$ the claim is proved.
	
	Let us show that $B$ is deep. Let $t\in\pl$, $j\in\N$ and $v\in [2^{j+1}-1,2^{j+2}-2]$. Since this guarantees all of $R\cap 2^{\log j'}$ is available from $B\harp v$, we have 
	\begin{equation}\label{equ.c}
	C^t(B\harp v) \geq C^n(B\harp v) \geq^+ C^{2^n}(r,v)
	\end{equation}
	where $r$ is the first string of size $\log j'$ in $R$. 
	Let  $t_1\in\ce$ be the time bound given by Theorem \ref{t.sym.inf} with $t_0(n)=2^n$, so that we have
	\begin{equation}\label{equ.b}
	C^{2^n}(r,v) \geq C^{t_1}(r) +C^{t_1}(v\ | \ r) -O(\log |v|)
	\end{equation}
	Let $p$ be testifying $C^{t_1}(v\ | \ r) $, i.e. $U(p,r)=v$ in at most $t_1(|v|)$ steps. Let $p'=\hat q\hat q'p$ where 
	$q$ are instructions, $q'$ is an encoding of $j$, 
	and $U$ on input $p'$:
      \begin{itemize}
         \item recovers $j$ then $j'$. 
         \item Computes $r$, i.e. the first element of $R[j'-1,2j'-2]$.
         \item simulates $U(p,r)=v$ (in $t_1(|v|)$ steps) and outputs $v$. 
      \end{itemize}
For the second item above: each  bit of $R$ requires at most $2^{\log j'}$ programs to be simulated for 
	$2^{2^{3\log j'}}$ steps, i.e. a total of $j'2^{j'^3}$ steps, thus a total of at most $2^{j'^{4}}\leq 2^{j^{5c}}$ steps to compute $R[j'-1,2j'-2]$. 
            The total running time is less than $t_2(|v|)=t_1(|v|)+2^{|v|^{5c}}$, since $|v|=1+ j$.
	Therefore we have
	\begin{equation}\label{equ.1}
	C^{t_2}(v)\leq C^{t_1}(v\ | \ r) +|\hat q| + |\hat q'| \leq^+ C^{t_1}(v\ | \ r) + 2\log j. 
	\end{equation}
        \noindent
	Also $B\in\cexp$ because $A\in \ce$ and $R[j'-1,2j'-2]$ requires at most $2^{j^{5c}}$ steps to be computed, i.e. $B\in\cexp$.

	Let $p$ testify $C^{t_2}(v)$, i.e. $U(p)=v$ in $t_2(|v|)$ steps.
	Let $p'=\hat qp$ where  
	$U$ on input $p'$ simulates $U(p)=v$ (in  $t_2(|v|)$ steps), and outputs $B\harp v$ (in $v2^{|v|^b}$ steps, where $B\in\text{Dtime}(2^{n^b})$) with a total running time less than
	$t_3(v)=t_2(|v|)+ v2^{|v|^b} \leq t_2(\log (v+1)) + v2^{\log^b (v+1)} \in 2^{\pl}$. Thus 
	\begin{equation}\label{equ.a}
	C^{t_3}(B\harp v) \leq^+ C^{t_2}(v). 
	\end{equation}
	We have,
	\begin{align*}
	&C^t(B\harp v) - C^{t_3}(B\harp v) \geq^+\\ 
      &C^{t_1}(r) +C^{t_1}(v\ | \ r) -O(\log |v|) - C^{t_2}(v) \geq \qquad &\text{By Equations \ref{equ.b}, and  \ref{equ.a}}\\
      &C^{t_1}(r) + C^{t_2}(v)- 2\log j-  C^{t_2}(v) -O( \log |v|)\geq \qquad &\text{By Equation \ref{equ.1}}\\
	&C^{t_1}(r)  -O(\log j)   
	\end{align*}

	Since $t_1\in\ce$, we have  $C^{t_1}(r)\geq |r|=\log j' \geq \log j^c$ because $r\in R$. Thus,
	$$C^t(B\harp v) - C^{t_3}(B\harp v) \geq c\log j -O(\log j) \geq 2 \log j \geq \log^{(2)} v $$
	for an appropriate choice of $c$. Since $t\in\pl$ and $ v\in\N$ are arbitrary, and $t_3\in 2^{\pl}$, $B$ is $(\log^{(2)}n)\pdeep$.
	\qed
	\end{proof}

\section{Lowness and Depth}
	The following result shows that some deep sets can be low. The idea of the proof is to construct a very sparse set of strings that are random at the polynomial time level but not at the exponential time level. The sparseness of the set guarantees that large queries can be answered with ``no'', hence only small queries need be computed, which guarantees lowness. To make the set deep, one needs to cut the sequence in blocks of subexponential size each containing a random string. The blocks need be large enough to not hurt the sparseness property, but small enough to ensure that the depth of the blocks is preserved over the whole  sequence.
	\begin{theorem}\label{t.low}
	For every $\epsilon > 0$, there exists a set in $\low{\ce}{\cexp}$ which is $\log^{1/(1+\epsilon)} n \pdeep$.
	\end{theorem}
	\begin{proof}
	Let $\epsilon> 0$, and $\epsilon' = \epsilon /2$.
	For all $n\geq 1$, define $T_{n+1}=2^{2^{\log^2 T_n}}$ with $T_0=1$. 
	Consider the following set $A$ where $A[T_n,T_{n+1}-1]$ is constructed as follows:
        For all  $k\leq \frac{4}{\log(1+\epsilon')}\log^{(2)}T_n$ with $2^{\log^{(1+\epsilon')^{\frac{k+1}{2}}}T_n}<T_{n+1}$, put the lex-first string $x^n_k$ of $R=\{x: \ C^{2^{n^2}}(x)\geq |x|\}$  with  
        $2^{\log^{(1+\epsilon')^{ \frac{k}{2}}}T_n} \leq \ind(x^n_k)< 2^{\log^{(1+\epsilon')^{ \frac{k+1}{2}}}T_n}$ 
        into $A$. 
        Note such a string exists since $\forall n\ R\cap\{0,1\}^n \neq\0$, and all the strings of length  $1 + \log^{(1+\epsilon')^{ \frac{k}{2}}}T_n$ are included in this interval: 
        strings of length $m$ have their index within $[2^m - 1, 2^{m+1} -2]$. Since $\log^{(1+\epsilon')^{ \frac{k+1}{2}}}T_n = (\log^{(1+\epsilon')^{ \frac{k}{2}}}T_n)^{(1+\epsilon')^{1/2}}$, the right endpoint of the interval is large enough, provided $T_n$ is large, because $(1+\epsilon')^{1/2}>1$. 
        Also add the lex-first string of $R$ of length $\log(T_{n+1})-1$. Note all strings of this length
	have their index within interval $[T_n,T_{n+1}-1]$.

	\begin{claim}
	$A\in\cexp$.
	\end{claim}
	This holds because to decide whether $x\in R$ one needs to simulate $2^{|x|}$ programs for $2^{|x|^{2}}$ steps i.e. less than $t_A(|x|)=2^{2|x|^{2}}$ steps.
	\begin{claim}
	$A$ is $\low{\ce}{\cexp}$. 
	\end{claim}
	Let us prove the claim. Let $L\in\ce^A$; we need to show that $L\in\cexp$.
	Let $M^A$ be an oracle TM deciding $L$ in time $2^{cn}$. Let $n\in\N$ and $v\in [T_n,T_{n+1}-1]$. $M^A(v)$ makes at most $2^{c|v|}$ queries of size at most $2^{c|v|}$ to $A$.
	First note that all queries are within $A\harp T_{n+2}$. Indeed since $|v|\leq \log T_{n+1}$, the largest query to $A$ has size at most $2^{c\log T_{n+1}}=T^c_{n+1}$, i.e. a string with index less than 
	$2^{1+T^c_{n+1}}< 2^{2^{\log^2 T_{n+1}}}=T_{n+2}$.
	Let us show the queries to $A$ can be answered in exponential time.
	\begin{subclaim}
	Let $q$ be a query to $A$ by $M^A(v)$ with  $|q|\geq |v|^4$. Then $C^{2cn}(q) < |q|$.
	\end{subclaim}
	Let us prove the claim. Any such $q$ can be specified by its index in $M^A(v)$'s queries list (i.e. $O(c|v|)$ bits) with the index of the previous queries to $A$ to be answered with yes (the other ones are answered with no),
        i.e. at most $O(|v| \log^2 T_n) < |v|^4$ bits because the max number of 1s (i.e. the number of yes queries) in $A[T_0,T_{n+2}-1]$ is less than 
	\begin{align*}
	a\log^{(2)}T_{n+1} +  a\log^{(2)}T_{n} +\ldots + a\log^{(2)}T_{1} &\leq a\log^{(2)}T_{n+1} +  an\log^{(2)}T_{n}\\
        &\leq (a+1)\log^{(2)}T_{n+1} \\
        &=(a+1)\log^2 T_n \\
        &\leq (a+1)|v|^2,
	\end{align*}
        because $|v|\geq \log T_n$, and where $a = \frac{4}{\log 1+\epsilon'}$.
	It takes at most $2^{c|v|}$ steps to simulate $M^A(v)$, plus $|v|^4$ steps to check the list of yes query answers, i.e. a total of less than $2^{2c|v|}$. 
	Thus $C^{2cn}(q)< |v|^4 \leq |q|$ which proves the claim. Consequently, if $|q|\geq |v|^4$, then $A(q)=0$ since $R(q)=0$. Hence only queries with $|q|< |v|^4$ need be computed to simulate $M^A(v)$.
	Such queries to $A$ are answerable in $2^{2|q|^2}\leq  2^{2|v|^8}$ steps, because $A\in\text{Dtime}(2^{2n^2})$. Thus $M^A(v)$ can be simulated in $2^{O(|v|^8)}$ steps, which proves $A$ is in $\low{\ce}{\cexp}$.

	\begin{claim}
        $A$ is $\log^{1/(1+\epsilon)}n \pdeep$.
	\end{claim}
        Let $t\in\pl, n \geq 1 $, and $v\in [T_n,T_{n+1}-1]$. Let $0<k<\frac{4}{\log(1+\epsilon')}\log^{(2)}T_n$ be such that $v\in [2^{\log^{(1+\epsilon')^{ \frac{k}{2}}}T_n},2^{\log^{(1+\epsilon')^{ \frac{k+1}{2}}}T_n}-1]$ ($k=0$ will be done later).
	Note that such a $k$ exists because when $k= \frac{4}{\log(1+\epsilon')}\log^{(2)}T_n$, then $2^{\log^{(1+\epsilon')^{ \frac{k+1}{2}}}T_n} > T_{n+1}$ (applying $\log$ on both sides of the equation three times).

	 We have 
	$$C^t(A\harp v) \geq C^n(A\harp v) \geq^+ C^{2^{n}}(x^n_{k-1},v)$$
	because given $A\harp v$ one can find $v$, and find $x^n_{k-1}$ (corresponding to the last bit equal to 1 in $A\harp 2^{\log^{(1+\epsilon')^{ \frac{k}{2}}}T_n} \prec A\harp v$) in $v\leq 2^{|v|}$ steps. 
	Let  $t_1\in\ce$ be the time bound given by choosing $t_0(n)=2^n$ in Theorem \ref{t.sym.inf}. By Theorem \ref{t.sym.inf}
    we have
	\begin{align}\label{eq.b}
	C^t(A\harp v) &\geq^+ C^{2^n}(x^n_{k-1},v) \geq C^{t_1}(x^n_{k-1}) + C^{t_1}(v\ | \ x^n_{k-1}) - O(\log|v|).  
	\end{align}

	Let $p$ be  a program testifying $C^{t_1}(v\ | \ x^n_{k-1})$, i.e. $U(p,x^n_{k-1})=v$ in $t_1(|v|)$ steps. 
	Let $p'=\hat q\hat q'p$ where $q$ are instructions, $q'$ encodes $k,n,\epsilon'$, and such that 
	$U(p')$ 
    \begin{itemize}
        \item recovers $k,n,\epsilon'$
        \item finds $x^n_{k-1}$ (i.e. the first string in $R$ with
            $2^{\log^{(1+\epsilon')^{ \frac{k-1}{2}}}T_n} \leq \ind(x^n_{k-1})< 2^{\log^{(1+\epsilon')^{ \frac{k}{2}}}T_n}$
    \item simulates $U(p,x^n_{k-1})=v$ 
    \item outputs $v$.
    \end{itemize}
	For each bit of  $R\harp x^n_{k-1} +1$, there are less than $2^{|x^n_{k-1}|}$ programs to simulate, each for at most $2^{|x^n_{k-1}|^2}$ steps, i.e. at most $2^{2|x^n_{k-1}|^2}\leq 2^{2|v|^2}$ steps. The simulation of $U(p,x^n_{k-1})=v$
	takes less than $t_1(|v|)$ steps, thus a total of $t_2(|v|)= t_1(|v|)+2^{2|v|^2}$ steps. Since $|\hat q'| \leq^+ 2\log k +2 \log n$, we have
	$$C^{t_2}(v)\leq^+ C^{t_1}(v\ | \ x^n_{k-1}) +2 \log k + 2 \log n < C^{t_1}(v\ | \ x^n_{k-1}) + \log |v| .$$

		Let $p'=\hat qp$ where  and $p$ is a minimal program testifying $C^{t_2}(v)$, and $q$ are instructions.
    $U$ on input $p'$ simulates $U(p)=v$, and outputs $A\harp v$. The simulation of $U(p)$ takes at most $t_2(|v|)$ steps. The computation of the first $v$ bits on $A$ takes time $v2^{2|s_v|^2}=v2^{2\log^2(v+1)}\leq 2^{\log^3 v}$ (because $A\in\text{Dtime}(2^{2n^2})$), i.e. a total of less than 
	$2^{\log^3 v}+t_2(|v|)= 2^{\log^3 v} + t_2(\log (v+1))$. Letting $t_3(n)= 2^{\log^3 n} + t_2(\log (n+1)) \in 2^{\pl}$, we have 
	$C^{t_2}(v) \geq^+ C^{t_3}(A\harp v)$, hence 
    \begin{equation}\label{eq.t3}
        C^{t_3}(A\harp v)\leq^+ C^{t_2}(v)   \leq C^{t_1}(v \ | \ x^n_{k-1}) + O(\log |v|). 
    \end{equation}

        We have,
	\begin{align*}
	&C^{t}(A\harp v) - C^{t_3}(A\harp v) \geq \\
        &C^{t_1}(x^n_{k-1}) +  C^{t_1}(v \ | \ x^n_{k-1})  - C^{t_1}(v \ | \ x^n_{k-1}) - O(\log |v|)\geq \quad &\text{By Equations \ref{eq.b} and \ref{eq.t3}}\\
	&|x^n_{k-1}| - O(\log |v|)\geq \quad &\text{because } x^n_{k-1}\in R\\
        &\log^{\frac{1}{1+\epsilon}} v
	\end{align*}
	because $|x^n_{k-1}| \geq \log^{(1+\epsilon')^{ \frac{k-1}{2}}} T_n$, thus 
    $|x^n_{k-1}|^{(1+\epsilon')} \geq \log^{(1+\epsilon')^{ \frac{k+1}{2}}} T_n
    > \log v$, thus 
    $|x^n_{k-1}| - O(\log |v|) > \log^{\frac{1}{1+\epsilon'}} v - O(\log |v|)
    = \log^{\frac{1}{1+\epsilon'}} v - O(\log^{(2)}v) 
    > \frac{1}{2} \log^{\frac{1}{1+\epsilon'}} v 
    > \log^{\frac{1}{1+\epsilon}} v$, for large enough $v$ and because $\epsilon' = \epsilon/2$.

	The case $k=0$ is similar: the same proof applies except that $x^n_{k-1}$ is replaced with the lex-last string $x$   whose index bit is 1 in $A[T_{n-1},T_n -1]$. By construction of $A$ the length of $x$ is $|x|=\log(T_n)-1$. Since $k=0$, 
    $\log v < \log^{(1+\epsilon')^{1/2}} T_n$, and similarly to the previous argument, we have
	\begin{align*}
        C^{t}(A\harp v) - C^{t_3}(A\harp v) &\geq 
        |x| - O(\log |v|) \\ 
        &\geq \log(T_n) - O(\log^{(2)} v) \\
        &\geq  \log^{\frac{1}{(1+\epsilon')^{1/2}}} v - O(\log^{(2)}v) \\
        &> \log^{\frac{1}{1+\epsilon'}} v \\
        &> \log^{\frac{1}{1+\epsilon}} v
	\end{align*}

    Since $v\in \N, t\in \pl$ are arbitrary and $t_3\in2^{\pl}$, $B$ is $\log^{1/(1+\epsilon)} n\pdeep$.
	\qed
	\end{proof}

\section{Discussion}
The slow growth law (Theorem \ref{t.sgl}) holds when both sets are in $\cexp$. It would be interesting to see whether this is required.

Although $\cexp$ contains a $\log n\pdeep$ set, the bounds in the slow growth law (Theorem \ref{t.sgl}) are not tight enough to prove that every $\cexp$-complete set is $\log n \pdeep$. It would be interesting to show that this holds (or not).
Highness and lowness have also been studied within the polynomial time hierarchy, i.e. for the class $\cnp$. Our polylog depth notion can be adapted to yield a depth measure between $\cp$ and $\cnp$; however, it is not obvious how to translate our highness and lowness results in that setting. Highness/lowness for $\ce$ is based on running time functions dominating each other
(similarly to the domination properties of high sets in computability theory). On the other hand,
highness for $\cnp$ is based on nested levels of nondeterministic computations (similarly to the arithmetical hierarchy in computability theory).

The techniques in Theorem \ref{t.low} fall short of constructing a polylog deep set that is low for E. It would be interesting to see if such a set exists.


\section*{Acknowledgement}
    We thank all the anonymous referees for useful comments on earlier versions of this paper.

\newcommand{\etalchar}[1]{$^{#1}$}

\end{document}